\documentclass[journal]{IEEEtran}
\usepackage[utf8]{inputenc}
\usepackage{newunicodechar}
\usepackage{graphicx}
\usepackage{subcaption}
\usepackage{cite}
\usepackage{amsmath,amssymb,multirow}
\usepackage{amsmath,amssymb}
\DeclareMathOperator{\E}{\mathbb{E}}

\usepackage{amsmath,amssymb}
\usepackage[noend]{algpseudocode}
\usepackage[linesnumbered,ruled,vlined]{algorithm2e}
\SetKwInOut{Input}{Input}
\SetKwInOut{Output}{Output\,}
\usepackage{adjustbox}
\usepackage{threeparttable}
\usepackage{upgreek}

\usepackage{textcomp}
\DeclareUnicodeCharacter{2212}{-}

\usepackage{graphicx}
\usepackage{textcomp}
\usepackage{xcolor}
\usepackage{color,soul}
\usepackage{dirtytalk}

\usepackage{scalerel}
\usepackage{tikz}
\usetikzlibrary{svg.path}
\usepackage[ruled,vlined]{algorithm2e}
\usetikzlibrary{svg.path}

\definecolor{orcidlogocol}{HTML}{A6CE39}
\tikzset{
  orcidlogo/.pic={
    \fill[orcidlogocol] svg{M256,128c0,70.7-57.3,128-128,128C57.3,256,0,198.7,0,128C0,57.3,57.3,0,128,0C198.7,0,256,57.3,256,128z};
    \fill[white] svg{M86.3,186.2H70.9V79.1h15.4v48.4V186.2z}
                 svg{M108.9,79.1h41.6c39.6,0,57,28.3,57,53.6c0,27.5-21.5,53.6-56.8,53.6h-41.8V79.1z M124.3,172.4h24.5c34.9,0,42.9-26.5,42.9-39.7c0-21.5-13.7-39.7-43.7-39.7h-23.7V172.4z}
                 svg{M88.7,56.8c0,5.5-4.5,10.1-10.1,10.1c-5.6,0-10.1-4.6-10.1-10.1c0-5.6,4.5-10.1,10.1-10.1C84.2,46.7,88.7,51.3,88.7,56.8z};
  }
}
\newcommand\orcidicon[1]{\href{https://orcid.org/#1}{\mbox{\scalerel*{
\begin{tikzpicture}[yscale=-1,transform shape]
\pic{orcidlogo};
\end{tikzpicture}
}{|}}}}

\usepackage{hyperref} 

\title{Multi-Fake Evolutionary Generative Adversarial Networks for Imbalance Hyperspectral Image Classification}

\author{\IEEEauthorblockN{Tanmoy Dam~$^{\orcidicon{0000-0003-3022-0971}}$\IEEEauthorrefmark{1}, Nidhi Swami\IEEEauthorrefmark{2},
Sreenatha G. Anavatti~$^{\orcidicon{; 0000-0002-4754-8191}}$\IEEEauthorrefmark{1},
Hussein A. Abbass,~\IEEEmembership{Fellow,~IEEE}~$^{\orcidicon{0000-0002-8837-0748}}$}\IEEEauthorrefmark{1}\\
\IEEEauthorblockA{\IEEEauthorrefmark{1}School of Engineering and Information Technology, University of New South Wales Canberra,  Australia.}
\IEEEauthorblockA{\IEEEauthorrefmark{2}Economics Department, Indian Institute of Technology Kharagpur,  India.}}

\begin{document}

\maketitle
\begin{abstract}
This paper presents a novel multi-fake evolutionary generative adversarial network(MFEGAN) for handling imbalance hyperspectral image classification. It is an end-to-end approach in which different generative objective losses are considered in the generator network to improve the classification performance of the discriminator network. Thus, the same discriminator network has been used as a standard classifier by embedding the classifier network on top of the discriminating function. The effectiveness of the proposed method has been validated through two hyperspectral spatial-spectral data sets. The same generative and discriminator architectures have been utilized with two different GAN objectives for a fair performance comparison with the proposed method. It is observed from the experimental validations that the proposed method  outperforms  the state-of-the-art-methods with better classification performance.
\end{abstract}

\begin{IEEEkeywords}
Imbalance multi-class classification, minority-oversampling, evolutionary GAN, hyperspectral datasets.
\end{IEEEkeywords}

\section{Introduction}
In hyperspectral imaging, instruments acquire images within many narrow, contiguous spectral bands throughout the spectrum that allows for acceptable discrimination between different features on the Earth’s surface. Several uses arise from collecting these data, such as land change detection, urban planning, classification, agriculture, identification, and surveillance \cite{guo2017learning}. New applications and technological challenges in data analysis arise from these earth observation data sources. Due to their high spectral resolution, many bands and abundant information mark hyperspectral images(HSI). These make the classification as the main challenging task.

The HSI classification has many applications, including land coverage detection, urban development planning, and managing resources under turbulence situations \cite{dam2020mixture}. However, the main challenges associated with the HSI classification task are (i) dimensionality and (ii) an imbalance in the number of training samples. A high degree of data redundancy and dimensionality can impede data analysis. Compared with other widely used supervised techniques like k-nearest-neighbors (KNN), neural network, and logistic regression, SVM-based classifiers usually perform well with limited training samples \cite{ghamisi2017advanced}. 

Considering the advancement of sensor and imaging systems, the spatial resolution of hyperspectral data is becoming increasingly fine. As a result, the classification performance can be significantly enhanced by using spatial information that can be extracted by filtering or segmentation approaches. Multiple kernel learning is commonly used for hyperspectral data classification due to the powerful capability to handle heterogeneous spectral
and spatial features efficiently \cite{ghamisi2017advanced}.

The spectral-spatial classification of hyperspectral data has been a subject of much research in recent years as HSI classification focuses largely on spatial-spectral approaches \cite{makki2017survey}. Sparse representations combined with Markov random fields(MRF)
are employed to explore spatial correlation that helps to improve classification performance \cite{yuan2015hyperspectral}. Discriminant analysis was applied to learn a representative subspace from the spectral-spatial domains that achieved good classification outcomes \cite{yuan2015hyperspectral, ghamisi2017advanced, feng2019classification}. Invariant and discriminatory features of the input data tend to be extracted by deep models, which typically contain two or more hidden layers \cite{zhong2019generative}. 
Previously\cite{ ghamisi2017advanced, feng2019classification,li2016hyperspectral}, the spectral characteristics of HSI were extracted using a deep-learning Convolutional Neural Network (CNN) which produced promising performance in classification. For example, there is a recently introduced method for extracting the features of HSI based on Gabor filtering and CNN, which leads to performance improvements \cite{hu2015deep}. In order to classify HSIs, one more framework based on principal components analysis (PCA), CNN and logistic regression were implemented. However, Deep CNN methods require a lot of balanced samples to train a large number of parameters, which is why in spite of great progress in HSI classification achieved by Deep learning models, they face the problem of over-fitting \cite{chen2017hyperspectral}. In addition, if the training data is not uniformly distributed (imbalanced problems) then the baseline classifier (CNN) suffers adequate performance due to major class biases. Generative adversarial network(GAN) is a popular methods due to its synthetic data generation capabilities. For an imbalanced learning paradigm, auxiliary classifier GAN (ACGAN) based method may improve the classification performance but it always suffers mode collapse issues due to major class biases \cite{hao2020annealing}. To overcome this, domain constraints three-players adversarial GAN game has been suggested where a class-dependent mixture of generators, a discriminator and a classifier are playing an adversarial game to improve the classifier performance \cite{dam2020mixture, dam2021does}. But this process requires class-dependent independent generators that forces the output to be within the data domain. Generated samples are always being selected by the classifier, which is relevant to it. This method was developed for spectral information rather than spatial resolution information, but spatial resolutions based generative models obtained better model accuracy \cite{ghamisi2017advanced}. 

In recent times, to tackle major class bias issues, evolutionary-based annealing genetic GAN (AGGAN) has been introduced to handle imbalance problems where simulated annealing (SA) approach is taken to get better classification performance of generated samples \cite{hao2020annealing}. The selection procedure is based on multiple generative losses in which the best losses will be survived during the learning procedure with the help of SA. However, to tackle the class-imbalance issue, minority oversampling methods were introduced to make the dataset in a balance form. The effectiveness of AGGAN has been validated through identically distributed image datasets. However, hyperspectral data contains spatial-spectral pieces of information. Hence, AGGAN may not be directly applicable over spatial-spectral HSI datasets to get adequate classification performance. To handle spatial-spectral HSIs datasets, we propose a minority over-sample MFEGAN method. The main contributions of this letter are described as follows,
\begin{itemize}
  \item We propose a minority over-sample MFEGAN for handling imbalance hyperspectral spatial-spectral image classification task.
  \item The proposed method has been validated on two hyperspectral datasets and compared with state-of-the-art methods including evolving ACGAN.
  \item To verify the statistical significance of the discriminator embedded classifier,  McNemar's test is carried out on all the methods.
\end{itemize}

\section{Background}
The supervised learning paradigm is broadly dealt with two approaches: discriminative and adversarial. The discriminative methods try to learn the discriminative features from the data directly.  A generative adversarial network is based on the adversarial games between two players (networks). The generative networks learn the parameters of real data distribution and generate real-like samples from the learned models. The learning of real distribution is achieved through either explicit or implicit methods under certain conditional assumptions. Finally, the learned parameters of generative networks can generate samples that mimic the actual distributions.  

 The GAN game has been developed on this unique learning strategy on the adversarial game principle \cite{goodfellow2014generative}. The GAN game is composed of two networks, Generator(G) and Discriminator(D). G model tries to capture all the distinctive modes in actual data distribution through a known distribution(e.g. normal). D network is working as a binary classifier between actual and generated samples. D network assigns a fake value if the samples are coming from the G network, whereas the actual data is considered a real value.The objective function of unsupervised GAN is defined as,
\begin{equation}
\label{eq:main}
\mathop {\min }\limits_G \mathop {\max }\limits_D \mathop J(G, D)  \E_{x \sim p_r} [logD(X_{real})] + \E_{z \sim p_g}[log(1-D(G(z)))]
\end{equation}
 where, $p_r$ is the real data distribution $(X_{real} \in p_r)$and $z \in p_g$ is the noise latent variable which takes from known prior to maps $X_{fake} = G(z)$ real like data distribution. 
 
\subsection{Proposed Approach}
The traditional GAN game can be extended to conditional GAN(CGAN) by considering class information($C$) in G and D networks since the class information($y \in C$) is embedded with latent prior(z) to generate the class-specific sample \cite{odena2017conditional}. Similarly, D networks also consider class information to control the generated space. By incorporating class conditionals($y \in C_i, i =1,2,...,N$), the generated samples are more likely to belong to any actual class distributions. The ACGAN was introduced by considering an auxiliary classifier on top of the D networks \cite{odena2017conditional}. The D network gives two outputs of probability distributions. One is associated with source data(real/fake), and another one is for the class score, indicted by ($P(S|X_{real/fake})$) and ($P(C|X_{real/fake})$), respectively. Finally, the final objective function is associated with two losses; source loss and classification loss.
\subsubsection{MFEGAN}
\begin{figure} 
 \includegraphics[keepaspectratio=true,scale=0.5]{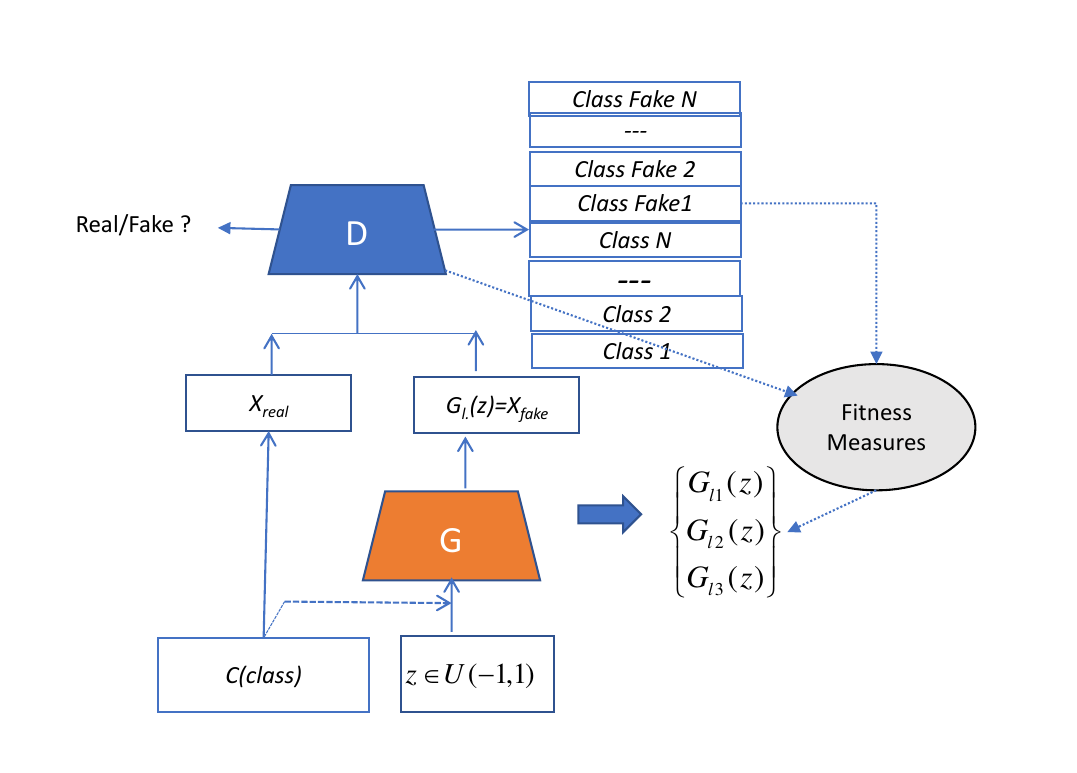}
 \caption{MFEGAN framework }
 \label{fig:MFEGAN_framework}
 \end{figure}

The proposed MFEGAN architecture is depicted in the Fig. \ref{fig:MFEGAN_framework}. The MFEGAN is an extension of ACGAN categories where evolutionary strategies have been considered to select the G. The background is similar to ACGAN where the discriminator classifies the real data (X) as $N[C_1, C_2,...,C_N]$ under the conditional class information $y$. But generated samples from evolving generator assigns as multiple fake classes $[C_{fake}1,C_{fake}2,..., C_{fake}i \in N ]$. However, in ACGAN, the $D$ assigns fake classes as part of real classes to $C_i \in N$.

The overall discriminator loss is associated with two losses: source loss($L^S$) and classification loss($L^C$). It is defined as follows,

\begin{align}
 L^D &= L^S +L^C  \notag  \\ 
     &= (E[\log P(S = real|X_{real})] + E[\log P(S = fake|X_{fake})]) \notag \\   
    & (E[\log P(C = C_i|X_{real})] + E[\log P(C=C_{fake}i|X_{fake})]) \notag \\
    &= (\E_{x \sim p_r} [log D(X,y)] + \E_{z \sim p_g}[log(1-D(G(z,y)))] \notag \\
    & (\E_{x \sim p_r} [log D^{real}(X,y)] + \E_{z \sim p_g}[log(1-D^{fake}(G(z,y)))]
 \label{equ:dis_update}
\end{align}
\subsubsection{Evolutionary strategy and Generator Updates}
In traditional ACGAN, the gradient of G is updated through the fixed D. In contrast to the traditional approach, the G network is updated through an evolutionary framework which is more suitable  to improve the diversity of the classifier. As a result, the classifier moves to the best optimum values. The evolutionary process follows two sub-stages: mutation and evolution(or selection). As described in \cite{wang2019evolutionary, hao2020annealing}, we have followed the same mutation process which are described as min-max mutation, heuristic-mutation and least-square mutation. These mutation techniques are nothing but each loss associated with different GAN methods. These mutation losses are defined as follows:

\begin{align}
   G_{l1}^{min-max} &= E[\log P(S = fake|X_{fake})]) \notag \\ 
                 & + (E[\log P(C = C_i|X_{fake})] \notag \\ 
               & = \E_{z \sim p_g}[log(1-D(G(z,y)))] \notag \\
            & + \E_{z \sim p_g}[log(1-D^{real}(G(z,y)))]
\label{l_min_max} 
\end{align}
\begin{align}
   G_{l2}^{heuristic} & = -E[\log P(S = fake|X_{fake})]) \notag \\ 
                   & + (E[\log P(C = C_i|X_{fake})] \notag \\ 
                   & = -\E_{z \sim p_g}[log(1-D(G(z,y)))] \notag \\
            & + \E_{z \sim p_g}[log(1-D^{real}(G(z,y)))]
\label{l_heuristic} 
\end{align}
\begin{align}
   G_{l3}^{least-square} & =  E[P(S = fake|X_{fake})]) \notag \\ 
                      & + (E[\log P(C = C_i|X_{fake})] \notag \\ 
                      & = \E_{z \sim p_g}[D(G(z,y))-1)^2] \notag \\
            & + \E_{z \sim p_g}[log(1-D^{real}(G(z,y)))]
\label{l_lease-square} 
\end{align}

All the three mutation losses are related to two parts: true/fake and multi-class classification. In multi-class classification, the generator is only updated through real class score($D^{real}$).

\textbf{Fitness measure:} 
To select the best generator losses during the learning procedure, it is required to have fitness on which the population(losses) will select. The fitness measure is based on two criteria: "generated samples belong to the real class" and "diversity between generated samples and the real class data". The generated samples belong to real class is represented by quality measure ($F_q$) as follows,
\begin{align}
   F_{q} = \{G_{l1}^{min-max},G_{l2}^{heuristic}, G_{l3}^{least-square} \}
\label{l_lease-square}
\end{align}
 The diversity measure($F_d$) is represented by gradient flow between real samples and generated samples through D networks and is defined as follows, 
\begin{align}
    F_d &= − \log || \nabla D − (F_q + E[\log P(S = real|X_{real})] \notag \\ 
    & + E[\log P(C = C_i|X_{real})]) ||
\end{align}

The final fitness measure is defined as follows,
\begin{align}
   F_{m} = \lambda \times F_q + F_d 
\label{fitness_measure}
\end{align}

Here $\lambda$ is the regulating factor, chosen as 0.5. Finally, one generator loss will  survive among three associated losses that has the maximum fitness value. 


\section{Experimental Validations}
\subsection{Design of G and D Architectures} 
G and D networks' architectures are listed in Table \ref{tab:Model_architecture}. G network consists of three deconvolutional (DeConv) layers followed by batch normalisation (BN) layer. To activate non-linearity in CNN structure, we have used Relu activation function. G takes each class conditionals uniform noise prior $(-1  to  1)$ to map each real class distributions. Both datasets are normalised within the limits of $0$ to $1$ . Sigmoid activation function is used in last layer of G networks. Similarly for D networks, we have used three  convolutional (Conv) layers followed by Leaky ReLU activation function. In addition, the D networks gives two classes as outputs: one is associated with true/fake and last one is dealt with class conditionals outputs. We have used two classes in the class-conditionals outputs, first one is associated with true classes and second one is used for fake classes.    
\begin{table}
\caption{G \& D Networks Details}
\begin{adjustbox}{width=\columnwidth,center}
    \begin{tabular}{|c|c|c|c|c|c|c|c|}
\hline 
Networks & Layer & Linear/DeConv/Conv & BN & Stride & Padding & Activation & Dropout\tabularnewline
\hline 
\multirow{4}{*}{G} & 1 & nn.Linear(100 +y, 512) & No & NA & NA & nn.ReLU() & No\tabularnewline
\cline{2-8} \cline{3-8} \cline{4-8} \cline{5-8} \cline{6-8} \cline{7-8} \cline{8-8} 
 & 2 & DeConv(512, 256, $SP/4$ $\times$ $SP/4$) & Yes & 1 & 0 & nn.ReLU() & No \tabularnewline
 \cline{2-8} \cline{3-8} \cline{4-8} \cline{5-8} \cline{6-8} \cline{7-8} \cline{8-8} 
 & 3 & DeConv(256, 128, 4 $\times$ 4) & Yes & 2 & 1 & nn.ReLU() & No \tabularnewline
 \cline{2-8} \cline{3-8} \cline{4-8} \cline{5-8} \cline{6-8} \cline{7-8} \cline{8-8} 
 & 4 & DeConv(128, 3, 4 $\times$ 4) & No & 2 & 1 & nn.Sigmoid() & No \tabularnewline
\hline 
\multirow{5}{*}{D} & 1 & Conv(3, 128, 4 $\times$ 4) & Yes  & 2 & 1 & LeakyReLU(0.2) & 0.5 \tabularnewline
\cline{2-8} \cline{3-8} \cline{4-8} \cline{5-8} \cline{6-8} \cline{7-8} \cline{8-8} 
 & 2 & Conv(128, 256, 4 $\times$ 4) & Yes & 2 & 1 & LeakyReLU(0.2) & 0.5 \tabularnewline
\cline{2-8} \cline{3-8} \cline{4-8} \cline{5-8} \cline{6-8} \cline{7-8} \cline{8-8} 
& 3 & Conv(256, 512, $SP/4$ $\times$ $SP/4$) & Yes & 1 & 0 & LeakyReLU(0.2) & 0.5 \tabularnewline
\cline{2-8} \cline{3-8} \cline{4-8} \cline{5-8} \cline{6-8} \cline{7-8} \cline{8-8} 
 & 4 & nn.Linear(512, 1) & NA & NA & NA & nn.Sigmoid() & No \tabularnewline
\cline{2-8} \cline{3-8} \cline{4-8} \cline{5-8} \cline{6-8} \cline{7-8} \cline{8-8} 
 & 4 & nn.Linear(512, y(real) + y(fake)) & NA & NA & NA & nn.LogSoftmax(dim =1) & No \tabularnewline
\hline 
   \end{tabular}
    \end{adjustbox}
\footnotesize{NA-Not Applicable, SP- Spatial Patches}
\label{tab:Model_architecture}
\end{table}

\subsection{Data Details and Experimental Validations}
Two popular hyperspectral datasets have been utilized to check the effectiveness of the proposed method. Indian Pines(IN) data was collected from Airborne Visible Infrared Imaging Spectrometer (AVIRIS) sensor over the Indian Pines test site at Northwestern Indiana regions. IN dataset contains $218$ bands of $145*145$ pixels of images in which water-absorbent bands were corrupted. 
After removal of water-absorbent bands, the final data consists of $200$ spectral bands. IN consists of $16$ landcover classes that that have spectral coverages from $0.4$ to $2.5 \upmu m $ with a spatial resolution of $20 m$. The $2nd$ dataset was collected at Kennedy space centre (KSC) Florida by using the NASA airborne AVIRIS instrument. The KSC dataset was collected with spatial resolutions of $18m$ throughout $20$kms. The captured dataset has a pixel size of $512 \times614$ with $13$ classes. Initially, the captured data has $224$ bands. Due to the water absorption, the fraction of bands with low SNR values is discarded from the original spectral bands. Finally, we have used $176$ bands with high SNR values for checking the classification performance.  Both the datasets are considered with spatial-spectral features which can lead us to give better classification performance \cite{ghamisi2017advanced}. We have taken the three major principle components for spectral features in which most of the data information is stored.  

\begin{figure} \includegraphics[keepaspectratio=true,scale=0.5]{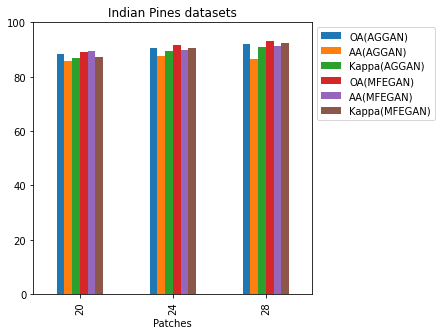}
 \caption{different patches of spatial domain}
 \label{fig: different patches performance}
 \end{figure}

\begin{figure}
\begin{subfigure}{.11\textwidth}
    \includegraphics[width=1\linewidth]{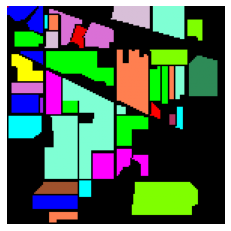}
  \caption{ground truth}
  \label{fig:sfig1}
\end{subfigure}%
   \begin{subfigure}{.11\textwidth}
    \includegraphics[width=1\linewidth]{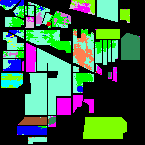}
  \caption{RF}
  \label{fig:sfig2}
\end{subfigure}
    \begin{subfigure}{.11\textwidth}
    \includegraphics[width=1\linewidth]{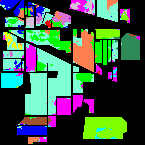}
  \caption{KNN}
  \label{fig:sfig3}
\end{subfigure}
    \begin{subfigure}{.11\textwidth}
    \includegraphics[width=1\linewidth]{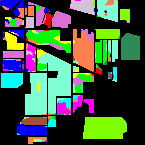}
  \caption{CNN}
  \label{fig:sfig4}
\end{subfigure}

\begin{subfigure}{.11\textwidth}
    \includegraphics[width=1\linewidth]{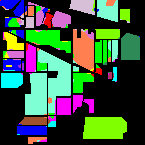}
  \caption{RO +CNN}
  \label{fig:sfig5}
\end{subfigure}
   \begin{subfigure}{.11\textwidth}
    \includegraphics[width=1\linewidth]{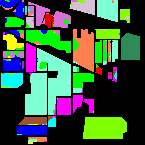}
  \caption{ACGAN}
  \label{fig:sfig6}
\end{subfigure}
  \begin{subfigure}{.11\textwidth}
   \includegraphics[width=1\linewidth]{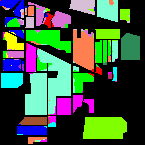}
  \caption{AGGAN}
  \label{fig:sfig7}
\end{subfigure}
   \begin{subfigure}{.11\textwidth}
    \includegraphics[width=1\linewidth]{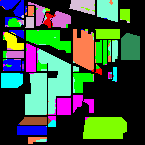}
  \caption{MFEGAN}
  \label{fig:sfig8}
\end{subfigure}
\caption{Classification maps for IN dataset in which each colour associated with each class. }
\label{fig:all_algorithm_performance IN}
\end{figure}

The whole datasets are divided into two parts in which training and testing samples are listed in Table \ref{Table: training and testing samples for IN & KSC datasets}. The imbalance ratio between major to minor class is obtained as ($\frac {123}{1})=123$ and $(\frac{100}{2})=50$ for IN and KSC datasets respectively. The total number of training and testing samples for both datasets are $512$, $9737$, $461$ and $4750$ respectively. Since the IN dataset is highly imbalanced,  we don't need to drop any of the classes to make dataset imbalance form. While training IN dataset, we have randomly chosen $5 \%$ to train the model and the remaining $ 95 \%$ samples were used to validate the model. The class distributions in the KSC dataset is almost balanced. To check the effectiveness of the proposed method, we have to make the dataset imbalanced in nature. Therefore, we have taken more samples from the $Water$ class with a factor of $50$ minor class (Swamp).
\begin{table}
\caption{Training \& Testing Samples for IN \& KSC datasets}
\resizebox{8cm}{!}{
\begin{tabular}{ccccc}
\hline 
Classes & \multicolumn{2}{c|}{IN} & \multicolumn{2}{c}{KSC}\tabularnewline
\cline{2-5} \cline{3-5} \cline{4-5} \cline{5-5} 
 & Training & Testing & Training & Testing\tabularnewline
\cline{2-5} \cline{3-5} \cline{4-5} \cline{5-5} 
0 & 2 & 44 & 33  & 728\tabularnewline
1 & 71  & 1357  & 23  & 220\tabularnewline
2 & 42 & 788  & 24  & 232\tabularnewline
3 & 12 & 225 & 24  & 228\tabularnewline
4 & 24  & 459 & 15  & 146\tabularnewline
5 & 36 & 694  &  22 & 207  \tabularnewline
6 & 1  & 27  & 2  & 103\tabularnewline
7 & 24  & 454  & 38  & 393\tabularnewline
8 & 1  & 19 & 51  & 469\tabularnewline
9 & 49  & 923 & 39 & 365 \tabularnewline
10 & 123  & 2332 & 41  & 378 \tabularnewline
11 & 30  & 563 & 49 & 454\tabularnewline
12 & 10  & 195 & 100 & 827\tabularnewline
13 & 63 & 1202 &  & \tabularnewline
14 & 19 & 367 &  & \tabularnewline
15 & 5  & 88 &  & \tabularnewline \\ \hline
{Sum}  & 512  & 9737 & 461 & 4750 \tabularnewline \\ \hline
\end{tabular}}
\label{Table: training and testing samples for IN & KSC datasets}
\end{table}
The multi-class classification performance of spatial-spectral hyperspectral datasets is compared with two standard machine learning methods such as KNN and RF, and deep generative frameworks. We have taken two popular auxiliary classifier-based generative models such as ACGAN \cite{odena2017conditional}, and AGGAN \cite{hao2020annealing}, for fair comparisons with the proposed method. We have also compared with the baseline CNN which is nothing but the D network working as a multi-class classification network. Similar to minority random oversampling  (RO) techniques used in ACGAN and AGGAN,  we have also compared MFEGAN with baseline D. The classification performance is observed with three popular measures in hyperspectral classification domains such as overall accuracy(OA), average accuracy (AA) and Kappa coefficients \cite{dam2020mixture}. In addition, each class classification performance is observed for both the datasets.
Table \ref{table:indian_pines_dataset} represents the classification performance for IN dataset. It is clearly observed from Table \ref{table:indian_pines_dataset} that MFEGAN obtained better performance among seven methods. This performance is based on $28$ spatial resolutions patches from the spatial domain data. The enhancement of performance indices is $1.3\%$ in OA, $5.61 \%$ in AA and $1.48 \%$ than the second-best results, AGGAN. The significant performance improvement is $10.52 \%$ OA, $13.59 \%$ in AA, and $11.94 \%$ from the baseline CNN. While dealing with imbalance classification problems, it is necessary to make the dataset balanced. Therefore, we have used randomly over-sample(RO) minority classes to balance and applied to baseline. It is clearly observed from the Table \ref{table:indian_pines_dataset} that $RO+CNN$
 obtained better performance improvement than the ACGAN by $1.72 \%$ in OA and  $2.0 \%$ in kappa except $0.04 \%$ in AA. 
\begin{table}
\caption{Qualitative performance on Indian Pines dataset}
\resizebox{9cm}{!}{
\begin{tabular}{cccccccc}
\hline 
\textbf{Method} & \textbf{RF} & \textbf{KNN} & \textbf{CNN} & \textbf{RO+CNN} & \textbf{ACGAN} & \textbf{AGGAN} & \textbf{MFEGAN}\tabularnewline
\hline 
\textbf{Alfalfa} & \textbf{100.00} & 66.67 & 81.81 & 48.10 & \textbf{90.47} & 58.73 & 83.78\tabularnewline
\textbf{Corn-notill} & 55.73  & 76.72 & \textbf{93.42} & 92.55 & 91.72 & 89.85 & 92.25\tabularnewline
\textbf{Corn-mintill} & 86.97 & 80.58 & 85.93 & 86.41 & 84.59 & 89.56 & \textbf{89.59}\tabularnewline
\textbf{Corn} & \textbf{100.00} & \textbf{100.00} & 46.40 & 80.76 & 66.79 & 91.59 & 97.10\tabularnewline
\textbf{Grass-pasture} & 98.63 & 81.44 & 76.71 & 93.13 & 92.41 & 97.42 & 98.00\tabularnewline
\textbf{Grass-trees} & 71.59  & 59.05 & \textbf{92.89} & 90.84 & 88.74 & 86.61 & 86.55\tabularnewline
\textbf{Grass-pasture-mowed} & 0.00  & 0.00 & 78.57 & \textbf{91.66} & 80.00 & 83.33 & 85.71\tabularnewline
\textbf{Hay-windrowed} & 96.59 & 96.05 & 85.17 & 98.48 & 98.05 & 97.63 & \textbf{99.12}\tabularnewline
\textbf{Oats} & 0.00 & 0.00 & 69.23 & 65.00 & 56.52 & 57.14 & \textbf{81.81}\tabularnewline
\textbf{Soybean-notill} & \textbf{97.00} & 73.14 & 73.96 & 87.75 & 74.82 & 90.89 & 94.41\tabularnewline
\textbf{Soybean-mintil} & 56.49 & 63.90 & 92.43 & 97.05 & \textbf{97.06} & 96.92 & 94.47\tabularnewline
\textbf{Soybean-clean} & \textbf{96.73}  & 93.01 & 62.79 & 74.30 & 82.27 & 83.74 & 88.78\tabularnewline
\textbf{Wheat} & \textbf{99.48} & 91.14 & 85.22 & 94.85 & 90.67 & 94.58 & 94.14\tabularnewline
\textbf{Woods} & 80.20  & 91.54 & \textbf{97.68} & 97.41 & 97.03 & 96.67 & 96.03\tabularnewline
\textbf{Buildings-Grass-Trees} & 94.01 & \textbf{97.66} & 83.16 & 77.60 & 77.07 & 82.78 & 93.83\tabularnewline
\textbf{Stone-Steel-Towers} & \textbf{100.00} & 96.00 & 80.64 & 80.23 & 93.33 & 85.18 & 84.70\tabularnewline
\hline 
\textbf{OA} & 69.61 & 74.70 & 84.33 & 90.69 & 90.35 & 92.02 & \textbf{93.22}\tabularnewline
\textbf{AA} & 77.09 & 72.96 & 80.38 & 84.76 & 86.72 & 86.41 & \textbf{91.26}\tabularnewline
\textbf{Kappa} & 64.04 & 70.57 & 82.24 & 89.43 & 88.97 & 90.91 & \textbf{92.26}\tabularnewline
\textbf{Time(epoch/sec)} & 1 & 1 & .98 & 1.16  & 2.55  & 8.03 & 6.96 \tabularnewline
\hline 
\end{tabular} \\
}
\footnotesize{Best obtained results marked as bold}\\
\label{table:indian_pines_dataset}
\end{table}
To obtain a better understanding of how spatial resolution patches have an impact on overall performance,  we have considered a set of three different spatial resolutions $\{ 20, 24, 28\}$ and compared with AGGAN. The illustration is depicted in Fig \ref{fig: different patches performance} in which all the three matrices are considered. It is clearly observed from Fig \ref{fig: different patches performance} that the proposed method outperformed  AGGAN in all the test cases,  in particular, the $28$-spatial resolution patches obtained the best performance among the three patches. 
Fig \ref{fig:all_algorithm_performance IN} represents classification performance for complete datasets using RF, KNN, CNN, RO+CNN, ACGAN, AGGAN, and MFEGAN. It is observed from the figure that MFEGAN performs well on the complete datasets. 

Furthermore, we have also considered the  KSC dataset for checking the effectiveness of the MFEGAN. Table \ref{tab:KSC_performance} describes the performance measures in terms of three parameters. Similar to IN dataset, our proposed method obtained better results among the seven methods. It is to be noted that the performance measures in the Table \ref{tab:KSC_performance} are for $28$ spatial resolution patches since, with this spatial size, we obtained the best performance for IN dataset. The enhancement in performance measures is $1.4 \%$ in OA, $2.1 \%$ in AA, and $1.6 \%$ in kappa compared to the AGGAN. Similarly, AGGAN is seen to have  better performance among the other state-of-the-art methods except for the proposed MFEGAN method. The baseline CNN gives the worst performance among all the methods due to the high imbalanced data distribution in the KSC dataset.  Besides, a significant improvement in baseline CNN performance is observed with the over-sampled minority classes in the datasets (RO+CNN). However, the baseline CNN could not reach the performance that was obtained  by the standard machine learning methods such as RF, KNN. The reason could be the over-fitting in the CNN parameters. Once, we over-sample minority classes, the performance improved a lot. Although, RO+CNN obtained better results compared to ACGAN but not AGGAN. In consistency for both datasets, AGGAN obtained $2nd$ best results compared to all the methods. It is also observed for both the cases, the AGGAN takes more time than the MFEGAN due to its SA block.  

\begin{table}
\caption{Qualitative performance on KSC dataset}
\resizebox{9cm}{!}{
\begin{tabular}{cccccccc}
\textbf{Classes} & \textbf{RF}  & \textbf{KNN} & \textbf{CNN} & \textbf{RO+CNN} & \textbf{ACGAN} & \textbf{AGGAN} & \textbf{MFEGAN}\tabularnewline
\hline 
\textbf{Scrub} & 75.83 & 75.40 & 76.02 & 91.83  & 93.52 & 93.89 & \textbf{96.30}\tabularnewline
\textbf{Willow swamp} & 90.79 & 90.80  & 08.63  & 89.68 & 93.33 & 92.76 & \textbf{98.59}\tabularnewline
\textbf{CP hammock} & 83.98 & 77.98  & 79.20 & 96.17 & 93.30 & 92.55 & \textbf{99.53}\tabularnewline
\textbf{Slash pine} & 87.38 & \textbf{98.90}  & 81.48 & 79.77 & 85.12 & 82.60 & 85.83\tabularnewline
\textbf{Oak}/\textbf{Broadleaf} & 99.20 & \textbf{100.00} & 79.38 & 95.94 & 93.37 & 99.23 & \textbf{100.00}\tabularnewline
\textbf{Hardwood} & 94.70  & 99.33 & 46.71  & 92.27 & 72.76 & 88.55 & \textbf{99.51}\tabularnewline
\textbf{Swamp} & 0.00 & 0. 00 & 80.00 & 95.87 & \textbf{100.00} & 95.09 & 95.29\tabularnewline
\textbf{Graminoid marsh} & 96.72  & 98.81 & 61.68  & 92.57 & 94.69 & \textbf{98.83} & 96.06\tabularnewline
\textbf{Spartina marsh} & \textbf{99.15} & 51.70 & 87.10  & 98.06 & 90.23  & 90.32 & 89.29\tabularnewline
\textbf{Cattail marsh} & 95.25 & 81.40 & 80.63 & 97.58 & 95.77  & \textbf{99.45} & 98.38 \tabularnewline
\textbf{Salt marsh} & \textbf{100.00} & \textbf{100. 00} & 96.89 & \textbf{100. 00} & \textbf{100.00} & \textbf{100.00} & \textbf{100. 00}\tabularnewline
\textbf{Mud flats} & 99.78 & \textbf{100.00} & 42.97  & 98.21 & 99.52 & \textbf{100.00} & \textbf{100. 00}\tabularnewline
\textbf{Water} & \textbf{100.00} & 97.63 & 04.76 & 99.63 & 96.61 & 97.64 & 98.10\tabularnewline
\hline 
\textbf{OA} & 92.54 & 82.63 & 58.80 & 95.15 & 93.49 & 95.22 & \textbf{96.58}\tabularnewline
\textbf{AA} & 86.37 & 82.46 & 63.50 & 94.43 & 92.94 & 94.69 & \textbf{96.68}\tabularnewline
\textbf{Kappa} & 91.65 & 80.47 & 55.30 & 94.60 & 92.75 & 94.67 & \textbf{96.19}\tabularnewline
Time(epoch/sec) & 1 & 1 & 0.98 & 1.24  & 5.13 & 17.56 & 15.71 \tabularnewline
\hline 
\end{tabular}
}
\footnotesize{Best obtained results marked as bold}\\
\label{tab:KSC_performance}
\end{table}
\begin{table}
\caption{McNemar's Tests($M_t$)}
\resizebox{.5\textwidth}{!}{
\begin{tabular}{lllllll}
\hline
\textbf{MFEGAN} & vs \textbf{RF} & vs \textbf{KNN} & vs \textbf{CNN} & vs \textbf{RO+CNN} & vs \textbf{ACGAN} & vs \textbf{AGGAN} \\ \hline
\textbf{IN}        & 38.20  & 32.25  & 18.49   & 6.21   & 6.97    & 3.08  \\ \hline
\textbf{KSC}      & 12.13  & 30.27 & 55.95    & 4.91      & 9.71  & 4.69 \\\hline
\end{tabular}}
\label{mcnemar_test}
\end{table}
The statistical significance test for both datasets has been discussed through McNemar’s test($M_t$) with the proposed method and state-of-the-art methods. This is provided in Table \ref{mcnemar_test}. Since the value of ($M_t>1.96(5\%)$), it indicates better statistical significance compared to the other methods.
\section{conclusion}
We have proposed a multi-fake evolutionary auxiliary classifier based GANs to improve the classification performance for IN and KSC datasets. For fair comparisons of the proposed method, other similar methods such as ACGAN, AGGAN and oversampling baseline methods have been considered. It is observed from the experimental validations that our proposed method outperformed all the methods while dealing with imbalanced datasets. Our future work will consider a mixture of two spatial-spectral generators for representing better feature representation learning for multi-class hyperspectral image classification. 
\bibliographystyle{IEEEtran}

\bibliography{references}

\begin{thebibliography}{10}
\providecommand{\url}[1]{#1}
\csname url@samestyle\endcsname
\providecommand{\newblock}{\relax}
\providecommand{\bibinfo}[2]{#2}
\providecommand{\BIBentrySTDinterwordspacing}{\spaceskip=0pt\relax}
\providecommand{\BIBentryALTinterwordstretchfactor}{4}
\providecommand{\BIBentryALTinterwordspacing}{\spaceskip=\fontdimen2\font plus
\BIBentryALTinterwordstretchfactor\fontdimen3\font minus
  \fontdimen4\font\relax}
\providecommand{\BIBforeignlanguage}[2]{{%
\expandafter\ifx\csname l@#1\endcsname\relax
\typeout{** WARNING: IEEEtran.bst: No hyphenation pattern has been}%
\typeout{** loaded for the language `#1'. Using the pattern for}%
\typeout{** the default language instead.}%
\else
\language=\csname l@#1\endcsname
\fi
#2}}
\providecommand{\BIBdecl}{\relax}
\BIBdecl

\bibitem{guo2017learning}
Y.~Guo, G.~Ding, L.~Liu, J.~Han, and L.~Shao, ``Learning to hash with optimized
  anchor embedding for scalable retrieval,'' \emph{IEEE Transactions on Image
  Processing}, vol.~26, no.~3, pp. 1344--1354, 2017.

\bibitem{dam2020mixture}
T.~Dam, S.~G. Anavatti, and H.~A. Abbass, ``Mixture of spectral generative
  adversarial networks for imbalanced hyperspectral image classification,''
  \emph{IEEE Geoscience and Remote Sensing Letters}, 2020.

\bibitem{ghamisi2017advanced}
P.~Ghamisi, J.~Plaza, Y.~Chen, J.~Li, and A.~J. Plaza, ``Advanced spectral
  classifiers for hyperspectral images: A review,'' \emph{IEEE Geoscience and
  Remote Sensing Magazine}, vol.~5, no.~1, pp. 8--32, 2017.

\bibitem{makki2017survey}
I.~Makki, R.~Younes, C.~Francis, T.~Bianchi, and M.~Zucchetti, ``A survey of
  landmine detection using hyperspectral imaging,'' \emph{ISPRS Journal of
  Photogrammetry and Remote Sensing}, vol. 124, pp. 40--53, 2017.

\bibitem{yuan2015hyperspectral}
Y.~Yuan, J.~Lin, and Q.~Wang, ``Hyperspectral image classification via
  multitask joint sparse representation and stepwise mrf optimization,''
  \emph{IEEE transactions on cybernetics}, vol.~46, no.~12, pp. 2966--2977,
  2015.

\bibitem{feng2019classification}
J.~Feng, H.~Yu, L.~Wang, X.~Cao, X.~Zhang, and L.~Jiao, ``Classification of
  hyperspectral images based on multiclass spatial--spectral generative
  adversarial networks,'' \emph{IEEE Transactions on Geoscience and Remote
  Sensing}, vol.~57, no.~8, pp. 5329--5343, 2019.

\bibitem{zhong2019generative}
Z.~Zhong, J.~Li, D.~A. Clausi, and A.~Wong, ``Generative adversarial networks
  and conditional random fields for hyperspectral image classification,''
  \emph{IEEE transactions on cybernetics}, 2019.

\bibitem{li2016hyperspectral}
W.~Li, G.~Wu, F.~Zhang, and Q.~Du, ``Hyperspectral image classification using
  deep pixel-pair features,'' \emph{IEEE Transactions on Geoscience and Remote
  Sensing}, vol.~55, no.~2, pp. 844--853, 2016.

\bibitem{hu2015deep}
W.~Hu, Y.~Huang, L.~Wei, F.~Zhang, and H.~Li, ``Deep convolutional neural
  networks for hyperspectral image classification,'' \emph{Journal of Sensors},
  vol. 2015, 2015.

\bibitem{chen2017hyperspectral}
Y.~Chen, L.~Zhu, P.~Ghamisi, X.~Jia, G.~Li, and L.~Tang, ``Hyperspectral images
  classification with gabor filtering and convolutional neural network,''
  \emph{IEEE Geoscience and Remote Sensing Letters}, vol.~14, no.~12, pp.
  2355--2359, 2017.

\bibitem{hao2020annealing}
J.~Hao, C.~Wang, H.~Zhang, and G.~Yang, ``Annealing genetic gan for minority
  oversampling,'' \emph{arXiv preprint arXiv:2008.01967}, 2020.

\bibitem{dam2021does}
T.~Dam, M.~M. Ferdaus, S.~G. Anavatti, S.~Jayavelu, and H.~A. Abbass, ``Does
  adversarial oversampling help us?'' \emph{arXiv preprint arXiv:2108.10697},
  2021.

\bibitem{goodfellow2014generative}
I.~Goodfellow, J.~Pouget-Abadie, M.~Mirza, B.~Xu, D.~Warde-Farley, S.~Ozair,
  A.~Courville, and Y.~Bengio, ``Generative adversarial nets,'' \emph{Advances
  in neural information processing systems}, vol.~27, 2014.

\bibitem{odena2017conditional}
A.~Odena, C.~Olah, and J.~Shlens, ``Conditional image synthesis with auxiliary
  classifier gans,'' in \emph{International conference on machine
  learning}.\hskip 1em plus 0.5em minus 0.4em\relax PMLR, 2017, pp. 2642--2651.

\bibitem{wang2019evolutionary}
C.~Wang, C.~Xu, X.~Yao, and D.~Tao, ``Evolutionary generative adversarial
  networks,'' \emph{IEEE Transactions on Evolutionary Computation}, vol.~23,
  no.~6, pp. 921--934, 2019.

\end{thebibliography}

\end{document}